\input harvmac
\def\ZZ{\hbox{Z\kern-.4emZ}}
\def\RR{\hbox{R\kern-.6emR}}

\lref\BredbergPV{
  I.~Bredberg, T.~Hartman, W.~Song and A.~Strominger,
  ``Black Hole Superradiance From Kerr/CFT,''
  arXiv:0907.3477 [hep-th].
}

\lref\LarsenXM{
  F.~Larsen,
  ``The Attractor Mechanism in Five Dimensions,''
Lect.\ Notes Phys.\  {\bf 755}, 249-281 (2008)
[hep-th/0608191].
}

\lref\CveticZQ{
  M.~Cveti\v c and C.~M.~Hull,
  ``Black holes and U-duality,''
  Nucl.\ Phys.\  B {\bf 480}, 296 (1996)
  [arXiv:hep-th/9606193].
}

\lref\BrownNW{
  J.~D.~Brown and M.~Henneaux,
  ``Central Charges in the Canonical Realization of Asymptotic Symmetries: An Example from Three-Dimensional Gravity,''
Commun.\ Math.\ Phys.\  {\bf 104}, 207-226 (1986).
}

\lref\MisnerKX{
  C.~W.~Misner,
  ``Interpretation of gravitational-wave observations,''
  Phys.\ Rev.\ Lett.\  {\bf 28}, 994 (1972).
}
\lref\TeukolskyHA{
  S.~A.~Teukolsky,
  ``Perturbations of a rotating black hole. 1. Fundamental equations for
  gravitational electromagnetic and neutrino field perturbations,''
  Astrophys.\ J.\  {\bf 185}, 635 (1973).
}

\lref\PressZZ{
  W.~H.~Press and S.~A.~Teukolsky,
  ``Perturbations of a Rotating Black Hole. II. Dynamical Stability of the Kerr
  Metric,''
  Astrophys.\ J.\  {\bf 185}, 649 (1973).
}

\lref\FrolovJH{
  V.~P.~Frolov and K.~S.~Thorne,
  Phys.\ Rev.\  D {\bf 39}, 2125 (1989).
}
\lref\CveticUW{
  M.~Cveti\v c and F.~Larsen,
  ``General rotating black holes in string theory: Greybody factors and  event
  horizons,''
  Phys.\ Rev.\  D {\bf 56}, 4994 (1997)
  [arXiv:hep-th/9705192].
}

\lref\CveticXV{
  M.~Cveti\v c and F.~Larsen,
  ``Greybody factors for rotating black holes in four dimensions,''
  Nucl.\ Phys.\  B {\bf 506}, 107 (1997)
  [arXiv:hep-th/9706071].
}

\lref\CveticVP{
  M.~Cveti\v c and F.~Larsen,
  ``Black hole horizons and the thermodynamics of strings,''
  Nucl.\ Phys.\ Proc.\ Suppl.\  {\bf 62}, 443 (1998)
  [Nucl.\ Phys.\ Proc.\ Suppl.\  {\bf 68}, 55 (1998)]
  [arXiv:hep-th/9708090].
}
\lref\CveticAP{
  M.~Cveti\v c and F.~Larsen,
  ``Greybody factors for black holes in four dimensions: Particles with
  spin,''
  Phys.\ Rev.\  D {\bf 57}, 6297 (1998)
  [arXiv:hep-th/9712118].
}

\lref\LarsenGE{
  F.~Larsen,
  ``A string model of black hole microstates,''
  Phys.\ Rev.\  D {\bf 56}, 1005 (1997)
  [arXiv:hep-th/9702153].
}

\lref\DiasNJ{
  O.~J.~C.~Dias, R.~Emparan and A.~Maccarrone,
  ``Microscopic Theory of Black Hole Superradiance,''
  Phys.\ Rev.\  D {\bf 77}, 064018 (2008)
  [arXiv:0712.0791 [hep-th]].
}

\lref\BredbergPV{
  I.~Bredberg, T.~Hartman, W.~Song and A.~Strominger,
  ``Black Hole Superradiance From Kerr/CFT,''
  arXiv:0907.3477 [hep-th].
}

\lref\StromingerSH{
  A.~Strominger and C.~Vafa,
  ``Microscopic Origin of the Bekenstein-Hawking Entropy,''
  Phys.\ Lett.\  B {\bf 379}, 99 (1996)
  [arXiv:hep-th/9601029].
}

\lref\CastroMS{
  A.~Castro, D.~Grumiller, F.~Larsen and R.~McNees,
  ``Holographic Description of AdS$_2$ Black Holes,''
  JHEP {\bf 0811}, 052 (2008)
  [arXiv:0809.4264 [hep-th]];
  A.~Castro and F.~Larsen,
  ``Near Extremal Kerr Entropy from AdS_2 Quantum Gravity,''
  arXiv:0908.1121 [hep-th].
}
\lref\HartmanDQ{
  T.~Hartman and A.~Strominger,
  ``Central Charge for AdS$_2$ Quantum Gravity,''
  JHEP {\bf 0904}, 026 (2009)
  [arXiv:0803.3621 [hep-th]].
}

\lref\GuicaMU{
  M.~Guica, T.~Hartman, W.~Song and A.~Strominger,
  ``The Kerr/CFT Correspondence,''
  arXiv:0809.4266 [hep-th].
}

\lref\SenQY{
  A.~Sen,
  ``Black Hole Entropy Function, Attractors and Precision Counting of
  Microstates,''
  Gen.\ Rel.\ Grav.\  {\bf 40}, 2249 (2008)
  [arXiv:0708.1270 [hep-th]].
}

\lref\BardeenPX{
  J.~M.~Bardeen and G.~T.~Horowitz,
  ``The extreme Kerr throat geometry: A vacuum analog of AdS(2) x S(2),''
  Phys.\ Rev.\  D {\bf 60}, 104030 (1999)
  [arXiv:hep-th/9905099].
}

\lref\BalasubramanianKQ{
  V.~Balasubramanian, A.~Naqvi and J.~Simon,
  ``A multi-boundary AdS orbifold and DLCQ holography: A universal  holographic
  description of extremal black hole horizons,''
  JHEP {\bf 0408}, 023 (2004)
  [arXiv:hep-th/0311237].
}
\lref\BalasubramanianBG{
  V.~Balasubramanian, J.~de Boer, M.~M.~Sheikh-Jabbari and J.~Simon,
  ``What is a chiral 2d CFT? And what does it have to do with extremal black
  holes?,''
  arXiv:0906.3272 [hep-th].
}

\lref\BarnichBF{
  G.~Barnich and G.~Compere,
  ``Surface charge algebra in gauge theories and thermodynamic integrability,''
  J.\ Math.\ Phys.\  {\bf 49}, 042901 (2008)
  [arXiv:0708.2378 [gr-qc]].
}
\lref\CompereAZ{
  G.~Compere,
  ``Symmetries and conservation laws in Lagrangian gauge theories with
  applications to the mechanics of black holes and to gravity in three
  dimensions,''
  arXiv:0708.3153 [hep-th].
}
\lref\BarnichKQ{
  G.~Barnich and G.~Compere,
  ``Conserved charges and thermodynamics of the spinning Goedel black hole,''
  Phys.\ Rev.\ Lett.\  {\bf 95}, 031302 (2005)
  [arXiv:hep-th/0501102].
}
\lref\BanadosDA{
  M.~Banados, G.~Barnich, G.~Compere and A.~Gomberoff,
  ``Three dimensional origin of Goedel spacetimes and black holes,''
  Phys.\ Rev.\  D {\bf 73}, 044006 (2006)
  [arXiv:hep-th/0512105].
}

\lref\EmparanEN{
  R.~Emparan and A.~Maccarrone,
  ``Statistical Description of Rotating Kaluza-Klein Black Holes,''
  Phys.\ Rev.\  D {\bf 75}, 084006 (2007)
  [arXiv:hep-th/0701150].
}

\lref\DiasNJ{
  O.~J.~C.~Dias, R.~Emparan and A.~Maccarrone,
  ``Microscopic Theory of Black Hole Superradiance,''
  Phys.\ Rev.\  D {\bf 77}, 064018 (2008)
  [arXiv:0712.0791 [hep-th]].
}

\lref\BardeenPX{
  J.~M.~Bardeen and G.~T.~Horowitz,
  ``The extreme Kerr throat geometry: A vacuum analog of AdS(2) x S(2),''
  Phys.\ Rev.\  D {\bf 60}, 104030 (1999)
  [arXiv:hep-th/9905099].
}

\lref\BalasubramanianRE{
  V.~Balasubramanian and P.~Kraus,
  ``A stress tensor for anti-de Sitter gravity,''
  Commun.\ Math.\ Phys.\  {\bf 208}, 413 (1999)
  [arXiv:hep-th/9902121].
}

\lref\SkenderisWP{
  K.~Skenderis,
  ``Lecture notes on holographic renormalization,''
  Class.\ Quant.\ Grav.\  {\bf 19}, 5849 (2002)
  [arXiv:hep-th/0209067].
}

\lref\AmselEV{
  A.~J.~Amsel, G.~T.~Horowitz, D.~Marolf and M.~M.~Roberts,
  ``No Dynamics in the Extremal Kerr Throat,''
  arXiv:0906.2376 [hep-th].
}
\lref\DiasEX{
  O.~J.~C.~Dias, H.~S.~Reall and J.~E.~Santos,
  ``Kerr-CFT and gravitational perturbations,''
  arXiv:0906.2380 [hep-th].
}

\lref\MaldacenaBW{
  J.~M.~Maldacena and A.~Strominger,
  ``AdS(3) black holes and a stringy exclusion principle,''
  JHEP {\bf 9812}, 005 (1998)
  [arXiv:hep-th/9804085].
}

\lref\DijkgraafFQ{
  R.~Dijkgraaf, J.~M.~Maldacena, G.~W.~Moore and E.~P.~Verlinde,
  ``A black hole farey tail,''
  arXiv:hep-th/0005003.
}

\lref\KrausVZ{
  P.~Kraus and F.~Larsen,
  ``Microscopic Black Hole Entropy in Theories with Higher Derivatives,''
  JHEP {\bf 0509}, 034 (2005)
  [arXiv:hep-th/0506176].
}

\lref\LarsenBU{
  F.~Larsen,
  ``Anti-de Sitter spaces and nonextreme black holes,''
  arXiv:hep-th/9806071.
}

\lref\CveticXV{
  M.~Cveti\v c and F.~Larsen,
  ``Greybody factors for rotating black holes in four dimensions,''
  Nucl.\ Phys.\  B {\bf 506}, 107 (1997)
  [arXiv:hep-th/9706071].
}

\lref\MaldacenaDS{
  J.~M.~Maldacena and L.~Susskind,
  ``D-branes and Fat Black Holes,''
  Nucl.\ Phys.\  B {\bf 475}, 679 (1996)
  [arXiv:hep-th/9604042].
}
\lref\KunduriVF{
  H.~K.~Kunduri, J.~Lucietti and H.~S.~Reall,
  ``Near-horizon symmetries of extremal black holes,''
  Class.\ Quant.\ Grav.\  {\bf 24}, 4169 (2007)
  [arXiv:0705.4214 [hep-th]].
}

\lref\StromingerYG{
  A.~Strominger,
  ``AdS(2) quantum gravity and string theory,''
  JHEP {\bf 9901}, 007 (1999)
  [arXiv:hep-th/9809027].
}
\lref\ChoFZ{
  J.~H.~Cho, T.~Lee and G.~W.~Semenoff,
  ``Two dimensional anti-de Sitter space and discrete light cone
  quantization,''
  Phys.\ Lett.\  B {\bf 468}, 52 (1999)
  [arXiv:hep-th/9906078].
}

\lref\GimonUR{
  E.~G.~Gimon and P.~Ho\v rava,
  ``Astrophysical Violations of the Kerr Bound as a Possible Signature of
  String Theory,''
  Phys.\ Lett.\  B {\bf 672}, 299 (2009)
  [arXiv:0706.2873 [hep-th]].
}
\lref\AzeyanagiBJ{
  T.~Azeyanagi, T.~Nishioka and T.~Takayanagi,
  ``Near Extremal Black Hole Entropy as Entanglement Entropy via AdS2/CFT1,''
  Phys.\ Rev.\  D {\bf 77}, 064005 (2008)
  [arXiv:0710.2956 [hep-th]].
}

\lref\GuptaKI{
  R.~K.~Gupta and A.~Sen,
  ``Ads(3)/CFT(2) to Ads(2)/CFT(1),''
  JHEP {\bf 0904}, 034 (2009)
  [arXiv:0806.0053 [hep-th]].
}

\lref\StromingerEQ{
  A.~Strominger,
  ``Black hole entropy from near horizon microstates,''
JHEP {\bf 9802}, 009 (1998)
[hep-th/9712251].
}

\lref\MaldacenaUZ{
  J.~M.~Maldacena, J.~Michelson and A.~Strominger,
  ``Anti-de Sitter fragmentation,''
  JHEP {\bf 9902}, 011 (1999)
  [arXiv:hep-th/9812073].
}

\lref\MaldacenaIH{
  J.~M.~Maldacena and A.~Strominger,
  ``Universal low-energy dynamics for rotating black holes,''
  Phys.\ Rev.\  D {\bf 56}, 4975 (1997)
  [arXiv:hep-th/9702015].
}
\lref\MathurET{
  S.~D.~Mathur,
  ``Absorption of angular momentum by black holes and D-branes,''
  Nucl.\ Phys.\  B {\bf 514}, 204 (1998)
  [arXiv:hep-th/9704156].
}
\lref\GubserQR{
  S.~S.~Gubser,
  ``Can the effective string see higher partial waves?,''
  Phys.\ Rev.\  D {\bf 56}, 4984 (1997)
  [arXiv:hep-th/9704195].
}

\lref\LarsenXM{
  F.~Larsen,
  ``The attractor mechanism in five dimensions,''
  Lect.\ Notes Phys.\  {\bf 755}, 249 (2008)
  [arXiv:hep-th/0608191].
}
\lref\CveticXZ{
  M.~Cveti\v c and D.~Youm,
  ``General Rotating Five Dimensional Black Holes of Toroidally Compactified
  Heterotic String,''
  Nucl.\ Phys.\  B {\bf 476}, 118 (1996)
  [arXiv:hep-th/9603100].
}

\lref\CveticKV{
  M.~Cveti\v c and D.~Youm,
  ``Entropy of Non-Extreme Charged Rotating Black Holes in String Theory,''
  Phys.\ Rev.\  D {\bf 54}, 2612 (1996)
  [arXiv:hep-th/9603147].
}

\lref\HartmanPB{
  T.~Hartman, K.~Murata, T.~Nishioka and A.~Strominger,
  ``CFT Duals for Extreme Black Holes,''
  JHEP {\bf 0904}, 019 (2009)
  [arXiv:0811.4393 [hep-th]].
}

\lref\LarsenGE{
  F.~Larsen,
  ``A String model of black hole microstates,''
Phys.\ Rev.\  {\bf D56}, 1005-1008 (1997)
[hep-th/9702153].
}

\lref\MaldacenaIX{
  J.~M.~Maldacena and A.~Strominger,
  ``Black hole greybody factors and D-brane spectroscopy,''
  Phys.\ Rev.\  D {\bf 55}, 861 (1997)
  [arXiv:hep-th/9609026].
}

\lref\MaldacenaIH{
  J.~M.~Maldacena and A.~Strominger,
  ``Universal low-energy dynamics for rotating black holes,''
  Phys.\ Rev.\  D {\bf 56}, 4975 (1997)
  [arXiv:hep-th/9702015].
}
\lref\CallanTV{
  C.~G.~.~Callan, S.~S.~Gubser, I.~R.~Klebanov and A.~A.~Tseytlin,
  ``Absorption of fixed scalars and the D-brane approach to black holes,''
  Nucl.\ Phys.\  B {\bf 489}, 65 (1997)
  [arXiv:hep-th/9610172].
}
\lref\GubserZP{
  S.~S.~Gubser and I.~R.~Klebanov,
  ``Four-dimensional greybody factors and the effective string,''
  Phys.\ Rev.\ Lett.\  {\bf 77}, 4491 (1996)
  [arXiv:hep-th/9609076].
}
\lref\ChowDP{
  D.~D.~K.~Chow, M.~Cveti\v c, H.~L\"u and C.~N.~Pope,
  ``Extremal Black Hole/CFT Correspondence in (Gauged) Supergravities,''
  arXiv:0812.2918 [hep-th].
}

\lref\LopesCardosoKY{
  G.~Lopes Cardoso, A.~Ceresole, G.~Dall'Agata, J.~M.~Oberreuter and J.~Perz,
 ``First-order flow equations for extremal black holes in very special
 geometry,''
  JHEP {\bf 0710}, 063 (2007)
  [arXiv:0706.3373 [hep-th]].
}

\lref\BalasubramanianEE{
  V.~Balasubramanian and F.~Larsen,
  ``Near horizon geometry and black holes in four-dimensions,''
Nucl.\ Phys.\  {\bf B528}, 229-237 (1998)
[hep-th/9802198].
}

\lref\HottaWZ{
  K.~Hotta and T.~Kubota,
  ``Exact Solutions and the Attractor Mechanism in Non-BPS Black Holes,''
  Prog.\ Theor.\ Phys.\  {\bf 118}, 969 (2007)
  [arXiv:0707.4554 [hep-th]].
}
\lref\GimonGK{
  E.~G.~Gimon, F.~Larsen and J.~Simon,
  ``Black Holes in Supergravity: the non-BPS Branch,''
  JHEP {\bf 0801}, 040 (2008)
  [arXiv:0710.4967 [hep-th]].
  ``Constituent Model of Extremal non-BPS Black Holes,''
  JHEP {\bf 0907}, 052 (2009)
  [arXiv:0903.0719 [hep-th]].
  }
\lref\BenaEV{
  I.~Bena, G.~Dall'Agata, S.~Giusto, C.~Ruef and N.~P.~Warner,
  ``Non-BPS Black Rings and Black Holes in Taub-NUT,''
  JHEP {\bf 0906}, 015 (2009)
  [arXiv:0902.4526 [hep-th]].
}
\lref\deBoerIP{
  J.~de Boer,
  ``Six-dimensional supergravity on S**3 x AdS(3) and 2d conformal field
  theory,''
  Nucl.\ Phys.\  B {\bf 548}, 139 (1999)
  [arXiv:hep-th/9806104].
}

\lref\KrausVZ{
  P.~Kraus, F.~Larsen,
  ``Microscopic black hole entropy in theories with higher derivatives,''
JHEP {\bf 0509}, 034 (2005).
[hep-th/0506176].
}

\lref\KastorGT{
  D.~Kastor and J.~H.~Traschen,
  ``A very effective string model?,''
  Phys.\ Rev.\  D {\bf 57}, 4862 (1998)
  [arXiv:hep-th/9707157].
}

\lref\DasWN{
  S.~R.~Das and S.~D.~Mathur,
  ``Comparing decay rates for black holes and D-branes,''
  Nucl.\ Phys.\  B {\bf 478}, 561 (1996)
  [arXiv:hep-th/9606185].
}

\lref\PeetES{
  A.~W.~Peet,
  ``The Bekenstein formula and string theory (N-brane theory),''
  Class.\ Quant.\ Grav.\  {\bf 15}, 3291 (1998)
  [arXiv:hep-th/9712253].
}
\lref\DavidWN{
  J.~R.~David, G.~Mandal and S.~R.~Wadia,
  ``Microscopic formulation of black holes in string theory,''
  Phys.\ Rept.\  {\bf 369}, 549 (2002)
  [arXiv:hep-th/0203048].
}
\lref\SenQY{
  A.~Sen,
  ``Black Hole Entropy Function, Attractors and Precision Counting of
  Microstates,''
  Gen.\ Rel.\ Grav.\  {\bf 40}, 2249 (2008)
  [arXiv:0708.1270 [hep-th]].
}

\lref\PiolinePF{
  B.~Pioline and J.~Troost,
  ``Schwinger pair production in AdS(2),''
  JHEP {\bf 0503}, 043 (2005)
  [arXiv:hep-th/0501169].
}
\lref\KimXV{
  S.~P.~Kim and D.~N.~Page,
  ``Schwinger Pair Production in $dS_2$ and $AdS_2$,''
  Phys.\ Rev.\  D {\bf 78}, 103517 (2008)
  [arXiv:0803.2555 [hep-th]].
}
\lref\AzeyanagiDK{
  T.~Azeyanagi, N.~Ogawa and S.~Terashima,
  ``The Kerr/CFT Correspondence and String Theory,''
  Phys.\ Rev.\  D {\bf 79}, 106009 (2009)
  [arXiv:0812.4883 [hep-th]].
}

\lref\BarnichJY{
  G.~Barnich and F.~Brandt,
  ``Covariant theory of asymptotic symmetries, conservation laws and  central
  charges,''
  Nucl.\ Phys.\  B {\bf 633}, 3 (2002)
  [arXiv:hep-th/0111246].
}
\lref\CompereIN{
  G.~Compere and S.~Detournay,
  ``Centrally extended symmetry algebra of asymptotically Goedel spacetimes,''
  JHEP {\bf 0703}, 098 (2007)
  [arXiv:hep-th/0701039].
}
\lref\CveticJA{
  M.~Cveti\v c and F.~Larsen,
  ``Statistical entropy of four-dimensional rotating black holes from
  near-horizon geometry,''
  Phys.\ Rev.\ Lett.\  {\bf 82}, 484 (1999)
  [arXiv:hep-th/9805146].
}
\lref\CveticXH{
  M.~Cveti\v c and F.~Larsen,
  ``Near horizon geometry of rotating black holes in five dimensions,''
  Nucl.\ Phys.\  B {\bf 531}, 239 (1998)
  [arXiv:hep-th/9805097].
}

\lref\ChongZX{
  Z.~W.~Chong, M.~Cveti\v c, H.~L\"u and C.~N.~Pope,
  ``Non-extremal rotating black holes in five-dimensional gauged
  supergravity,''
  Phys.\ Lett.\  B {\bf 644}, 192 (2007)
  [arXiv:hep-th/0606213].
}

\lref\CveticJN{
  M.~Cveti\v c and F.~Larsen,
  ``Greybody Factors and Charges in Kerr/CFT,''
  JHEP {\bf 0909}, 088 (2009)
  [arXiv:0908.1136 [hep-th]].
}

\lref\GuicaMU{
  M.~Guica, T.~Hartman, W.~Song and A.~Strominginer,
  ``The Kerr/CFT Correspondence,''
Phys.\ Rev.\  {\bf D80}, 124008 (2009)
[arXiv:0809.4266 [hep-th]].
}

\lref\CastroFD{
  A.~Castro, A.~Maloney and A.~Strominger,
  ``Hidden Conformal Symmetry of the Kerr Black Hole,''
  Phys.\ Rev.\  D {\bf 82}, 024008 (2010)
  [arXiv:1004.0996 [hep-th]].
}

\lref\MaldacenaIX{
  J.~M.~Maldacena and A.~Strominger,
  ``Black hole grey body factors and d-brane spectroscopy,''
Phys.\ Rev.\  {\bf D55}, 861-870 (1997)
[hep-th/9609026].
}

\lref\BredbergHP{
  I.~Bredberg, C.~Keeler, V.~Lysov and A.~Strominger,
  ``Cargese Lectures on the Kerr/CFT Correspondence,''
[arXiv:1103.2355 [hep-th]].
}

\lref\CveticMN{
  M.~Cveti\v c, G.~W.~Gibbons and C.~N.~Pope,
  ``Universal Area Product Formulae for Rotating and Charged Black Holes in Four and Higher Dimensions,''
  Phys.\ Rev.\ Lett.\  {\bf 106}, 121301 (2011)
[arXiv:1011.0008 [hep-th]].
}

\lref\MaldacenaDS{
  J.~M.~Maldacena and L.~Susskind,
  ``D-branes and fat black holes,''
  Nucl.\ Phys.\  B {\bf 475}, 679 (1996)
  [arXiv:hep-th/9604042].
}


\lref\KrishnanPV{
  C.~Krishnan,
  ``Hidden Conformal Symmetries of Five-Dimensional Black Holes,''
JHEP {\bf 1007}, 039 (2010)
[arXiv:1004.3537 [hep-th]];
  D.~Chen, P.~Wang and  H.~Wu,
  ``Hidden conformal symmetry of rotating charged black holes,''
Gen.\ Rel.\ Grav.\  {\bf 43}, 181-190 (2011)
[arXiv:1005.1404 [gr-qc]];
  M.~Becker, S.~Cremonini and W.~Schulgin,
  ``Correlation Functions and Hidden Conformal Symmetry of Kerr Black Holes,''
JHEP {\bf 1009}, 022 (2010)
[arXiv:1005.3571 [hep-th]];
  H.~Wang, D.~Chen, B.~Mu and H.~Wu,
  ``Hidden conformal symmetry of extreme and non-extreme Einstein-Maxwell-Dilaton-Axion black holes,''
JHEP {\bf 1011}, 002 (2010)
[arXiv:1006.0439 [gr-qc]];
  C.~-M.~Chen, Y.~-M.~Huang, J.~-R.~Sun, M.~-F.~Wu and S.~-J.~Zou,
  ``On Holographic Dual of the Dyonic Reissner-Nordstrom Black Hole,''
Phys.\ Rev.\  {\bf D82}, 066003 (2010)
[arXiv:1006.4092 [hep-th]];
  I.~Agullo, J.~Navarro-Salas, G.~J.~Olmo and L.~Parker,
  ``Hawking radiation by Kerr black holes and conformal symmetry,''
Phys.\ Rev.\ Lett.\  {\bf 105}, 211305 (2010)
[arXiv:1006.4404 [hep-th]];
  K.~-N.~Shao and Z.~Zhang,
  ``Hidden Conformal Symmetry of Rotating Black Hole with four Charges,''
Phys.\ Rev.\  {\bf D83}, 106008 (2011)
[arXiv:1008.0585 [hep-th]];
  A.~M.~Ghezelbash, V.~Kamali and M.~R.~Setare,
  ``Hidden Conformal Symmetry of Kerr-Bolt Spacetimes,''
Phys.\ Rev.\  {\bf D82}, 124051 (2010)
[arXiv:1008.2189 [hep-th]];
  D.~A.~Lowe, I.~Messamah and A.~Skanata,
  ``Scaling dimensions in hidden Kerr/CFT,''
[arXiv:1105.2035 [hep-th]];
  S.~Bertini, S.~L.~Cacciatori and  D.~Klemm,
  ``Conformal structure of the Schwarzschild black hole,''
[arXiv:1106.0999 [hep-th]].
}

\lref\MaldacenaDR{
  J.~M.~Maldacena and  L.~Maoz,
  ``Desingularization by rotation,''
JHEP {\bf 0212}, 055 (2002)
[hep-th/0012025].
}
\lref\LuninIZ{
  O.~Lunin, J.~M.~Maldacena and L.~Maoz,
  ``Gravity solutions for the D1-D5 system with angular momentum,''
[hep-th/0212210].
}
\lref\BalasubramanianRT{
  V.~Balasubramanian, J.~de Boer, E.~Keski-Vakkuri and S.~F.~Ross,
  ``Supersymmetric conical defects: Towards a string theoretic description of black hole formation,''
Phys.\ Rev.\  {\bf D64}, 064011 (2001)
[hep-th/0011217].
}
\lref\DijkgraafFQ{
  R.~Dijkgraaf, J.~M.~Maldacena, G.~W.~Moore and  E.~P.~Verlinde,
  ``A Black hole Farey tail,''
[hep-th/0005003].
}

\Title{\vbox{\baselineskip12pt 
\hbox{UPR-1228-T} 
\vskip-.5in}
}
{\vbox{\centerline{Conformal Symmetry for General Black Holes}
}}
\medskip
\centerline{\it
Mirjam Cveti\v c${}^{1,2}$ and Finn Larsen${}^{3}$
}
\bigskip
\centerline{${}^1$Department of Physics and Astronomy, University of Pennsylvania, Philadelphia, PA-19104, USA.}
\smallskip
\centerline{${}^2$Center for Applied Mathematics and Theoretical Physics,
University of Maribor, Maribor, Slovenia.}
\smallskip
\centerline{${}^3$Michigan Center for Theoretical Physics, 450 Church St., Ann Arbor,
MI-48109, USA.}
\smallskip

\vglue .3cm
\bigskip\bigskip\bigskip
\centerline{\bf Abstract}
\noindent

We show that the warp factor of a generic asymptotically flat
black hole in five dimensions
can be adjusted such that a conformal symmetry emerges. The construction preserves all
near horizon properties of the black holes, such as the thermodynamic potentials and 
the entropy. We interpret the geometry with modified asymptotic behavior 
as the ``bare" black hole, with the ambient flat space removed. 
Our warp factor subtraction generalizes hidden conformal symmetry and applies whether or not rotation is significant. We also find a relation to standard
AdS/CFT correspondence by embedding the black holes in six
dimensions.
The asymptotic conformal symmetry guarantees a dual CFT description 
of the general rotating black holes.

\Date{}

\newsec{Introduction}
Ever since the early days of string theory it has been speculated that a 2D CFT 
could be responsible for the microscopic interpretation of black hole entropy.
The advent of precise realizations of this vision in the context supersymmetric
black holes has made the speculations even more appealing, also in settings
far from extremality. Indeed, in the string theory community it is widely assumed 
that the black hole entropy can be accounted for quite generally, in much the same 
way as it has been for supersymmetric black holes; and this purportedly responds 
to Hawking's challenge to quantum mechanics, embodied in the information
loss paradigm. However, despite the optimistic conventional wisdom, little 
concrete progress has been towards a CFT interpretation of black holes far from 
extremality. 

One of the challenges faced by any attempt to make a CFT interpretation precise
for general black holes is that generic black holes have negative specific 
heat. For example, this is the case for most Kerr black holes, 
including Schwarzchild black holes. This feature of black hole thermodynamics reflects the physical coupling 
between the internal structure of the black hole and modes that escape to infinity. 
In order to focus on the black hole ``by itself" one must necessarily imagine enclosing 
the black hole in a box that reflects the emanating radiation and returns it to the black 
hole, thus creating an equilibrium system. This complication must be taken into 
account in any precise discussion of black hole thermodynamics, but its necessity 
is especially imposing if one seeks a dual CFT description, since unitary CFTs 
always have positive specific heat. In this paper we respond to this necessity in 
a manner that incorporates several other attractive ideas. 

An apparently unrelated clue to the internal structure of black holes involves
a massless scalar field probing a general black hole background. It was noticed 
a long time ago that the wave equation in this setting has remarkable simplifications 
even for general
black holes \CveticUW. In particular there is an $SL(2,\RR)^2$
symmetry, when certain terms are removed. The offending terms are indeed negligible
in many special cases, including the near extreme limit (the AdS/CFT correspondance) 
\refs{\MaldacenaIX,\CveticUW}, the near extreme rotating limit (the Kerr/CFT correspondance) 
\refs{\GuicaMU,\CveticJN}, and the 
low energy limit \refs{\CveticUW,\CastroFD}. However, in general there is no $SL(2,\RR)^2$
symmetry, just like not all black holes geometries have a near horizon AdS$_3$
component. This would seem to doom a CFT interpretation of the general case. 
However, the recent proposal dubbed ``hidden conformal symmetry'' asserts that
the conformal symmetry suggested by the massless wave equation is useful 
generally after all \CastroFD --- it is just that it is spontaneously broken. 
This approach has been developed by many researchers, 
including \refs{\KrishnanPV}.

The main technical result of this paper is that we construct the geometry corresponding
to the wave equation exhibiting $SL(2,\RR)^2$ symmetry.  In other words, we find the geometrical counterpart to the omission of terms violating $SL(2,\RR)^2$ in the wave 
equation. We refer to the resulting geometry as the ``subtracted geometry'', since it 
corresponds to removing certain terms in an overall warp factor. 
The physical interpretation we propose for the subtraction procedure
is that it corresponds to enclosure of the black hole in a box: it is the 
asymptotic Minkowski space that cannot be attributed to the black hole
that is being subtracted. 

As we have noted, a box delimiting the black hole from its surroundings is inevitable if we 
seek to identify a dual CFT. The added value offered by the specific box we construct is 
that it preserves conformal invariance and it is consistent with separation of variables. 
The subtracted geometry has the same thermodynamic 
potentials and entropy as the full geometry,
and it employes the same time. The only part of the geometry that has been changed is a
certain warp factor, and that only in its asymptotic behavior. 

The wave equation for a massless scalar field probing the subtracted geometry exhibits 
$SL(2,\RR)^2$ symmetry, by construction. However, the geometrical interpretation of
this symmetry remains 
obscure {\it a priori}. Progress can be made by lifting the geometry with subtracted conformal
factor from 5D to 6D. Indeed, the subtracted geometry is recognized 
after the lift as locally AdS$_3\times S^3$ geometry. Moreover,  
the global identifications are such that the AdS$_3$ factor is precisely the BTZ black hole;
and the $S^3$ is fibered over the AdS$_3$ in the manner familiar from rotating
black holes near extremality \CveticXH. We stress again that, here, we identify these features
in the geometry of black holes that are generally far from extremality. 

In this paper we focus on the ``mesoscopic'' analysis of black holes, ie. we seek to infer
features of the microscopic theory from the classical geometry. 
However, the structure that 
we pursue may persist in the full quantum theory. Given that supersymmetric 
black holes are described in detail by purely holomorphic 
CFTs with large central charge we expect that other black holes are 
similarly described by CFTs with large central charge, albeit no longer holomorphic ones. The semi-classical
level matching condition that applies to the black hole entropy all the way off extremality
is encouraging for this program \LarsenGE, as is the (related) semiclassical quantization condition
on the areas of the black hole horizons \refs{\CveticUW,\CveticMN}. 

This paper is organized as follows. In section 2 we review the general 5D black hole 
in string theory, with three charges and two angular momenta. Specifically, we need
the representation of the metric as a 4D base with time appearing as a $U(1)$ fibration. We derive the
thermodynamic potentials for a large family of geometries taking this form. 
In section 3 we present the subtracted metric and the wave equation in this background. These are the key technical results. In section 4, we rewrite 
the subtracted 5D metric in a 6D form, by introducing an auxiliary 
coordinate. This gives a linear realization of the conformal symmetry. 
In section 5 we compare our explicit construction to the hidden conformal symmetry program. 

\newsec{General 5D Black Holes in String Theory}
In this section we review the canonical family of asymptotically
flat string theory black holes in $D=5$ spacetime 
dimensions \CveticXZ. These black holes are the most general solutions in $N=4,8$ string 
theory, up to duality transformations on the matter sector \CveticZQ. 
We 
employ the recently uncovered form of 
the metric as a 4D base with time represented as a $U(1)$ fibration \ChongZX.

We also derive thermodynamic potentials and other parameters directly
from the geometry. This computation will be organized in order that it 
guide the subsequent construction of a suitable ``box" for these 
black holes. 

\subsec{The Geometry}
The independent parameters of the black hole are the 
mass, two angular momenta, and  three charges. 
They are parametrized as
\eqn\cg{\eqalign{
{4G_5\over\pi}M &= {1\over 2} \mu\sum_{i=1}^3\cosh 2\delta_i~,\cr
{4G_5\over\pi}Q_i & = {1\over 2} \mu \sinh 2\delta_i~,~~~ (i=1,2,3)~,\cr
{4G_5\over\pi}J_{R,L} & =  {1\over 2} \mu (b\pm a) \left( \Pi_c \mp \Pi_s\right)~,
}}
where
\eqn\cga{
\Pi_c \equiv \prod_{i=1}^3\cosh\delta_i 
~,~~~ \Pi_s \equiv  \prod_{i=1}^3 \sinh\delta_i~.
}

We write the 5D metric as a fibration over a 4D base space \ChongZX
\eqn\cb{\eqalign{
ds^2_5 & = - \Delta^{-2/3}_0 G ( dt+{\cal A})^2 + \Delta^{1/3}_0 ds^2_4~,\cr
ds^2_4 &= {dx^2\over 4X} + {dy^2\over 4Y} + 
{U\over G} (d\chi - {Z\over U} d\sigma)^2 + {XY\over U} d\sigma^2~,
}}
where for the black holes we consider
\eqn\cd{\eqalign{
X & = (x+a^2) (x+b^2) - \mu x~, \cr
Y & = - (a^2 - y) (b^2 - y)~,\cr
\Delta_0 & = (x+y)^3 H_1 H_2 H_3~, \cr
H_i & = 1 + {\mu \sinh^2\delta_i\over x+y}~,~(i=1,2,3)~,\cr
G & = (x+y) (x+y - \mu )~, \cr
{\cal A} & = {\mu \Pi_c \over x + y - \mu}[ (a^2 + b^2 - y)d\sigma - abd\chi]
-{\mu  \Pi_s \over x+y} (abd\sigma - y d\chi)~,\cr
U & = yX - xY~, \cr
Z & = ab(X+Y)~.
}}

The base space coordinates $(x,y,\sigma,\chi)$ are 
related to the more familiar radial coordinates as
\eqn\ce{\eqalign{
x& = r^2~,\cr
y&=a^2 \cos^2\theta + b^2\sin^2\theta~,\cr
\sigma & = {1\over a^2- b^2} \left(a\phi
-b\psi \right)~,\cr
\chi & = {1\over a^2- b^2} \left( b\phi
-a\psi\right)~.
}}
The $(r,\theta,\phi,\psi)$ coordinates are such that the base 
metric asymptotically approaches flat space in the conventional form
\eqn\cf{\eqalign{
ds^2_4 &\sim dr^2 + r^2 d\Omega^2_3~,\cr
d\Omega^2_3 &= d\theta^2 + \sin^2\theta d\phi^2 + \cos^2\theta d\psi^2~.
}}
We assume $b^2>a^2$ so that $y\in (a^2,b^2)$ as $\theta\in (0,\pi/2)$. 

The advantage of employing the radial coordinate $x$ and the polar coordinate $y$, rather than the
more conventional $r$, $\theta$, is that with this parametrization the 
radial and polar coordinates appear 
in a roughly symmetric manner. The azimuthal coordinates $\sigma$, $\chi$ parametrize the angular 
isometries.  

\subsec{Black Hole Thermodynamics}
\noindent
The geometry \cb\ is expressed in terms of a large number of functions:
\item
~i) 
The radial function $X$ depends only on the radial coordinate $x$.
\item
~ii)
The
azimuthal function $Y$ depends on the azimuthal coordinate $y$.
\item
~iii)
The warp-factors $\Delta_0$ and $G$ depend only on the combination
$x+y$. 
\item
~iv)
The remaining functions $U, Z$ and the one-form ${\cal A}$ depend on 
the coordinates $x, y$ independently. 

\noindent
An instructive way to appreciate the geometry is to work out the thermodynamics of black holes taking the 
general form \cb. In other words, we assume the 
dependences on $x, y$ given above, 
but we will not employ the particular functions \cd. 

None of the coordinates $t, \chi, \sigma$ appear in the metric function,
so they all parametrize isometries. 
The coordinate $t$ is special because it represents time in the 
asymptotic spacetime. Accordingly, the {\it static limit} is the surface 
\eqn\cf{
G=0~,
}
, since this is where $g_{tt}=0$. As this surface is 
crossed, trajectories along the coordinate $t$ cease to be time-like.
The {\it ergosphere} is the volume inside this surface (but outside the 
event horizon.) 

In the ergosphere, physical (ie. light-like) trajectories at fixed $x, y$ always have a component along one or both of the azimuthal angles $\chi,\sigma$. 
Physically this means they must co-rotate along with the rotating black 
hole. A light-like
combination of $t, \chi, \sigma $ can be found as long as the 
sub-determinant in the $t-\chi-\sigma$ space
\eqn\cg{
{\rm det} ~g(t-\chi-\sigma) = - XY~,
}
remains negative. Since $Y>0$ (except at the poles\foot{We keep the polar angle $y$ fixed in most computations. 
We also
do not analyze the poles at $Y=0$ which are quite singular in the present coordinates.}), 
this condition identifies the {\it event horizon} as the (outer component of the) locus
\eqn\ch{
X=0~.
}
On the inner side of this surface $X<0$, so there all linear combinations 
of $t, \chi, \sigma$ are spacelike. Then physical trajectories must move radially along $x$: capture by the black hole has become inevitable.

Outside the event horizon we can identify a notion of time at any 
given $x,y$ by diagonalizing the metric in the $(t, \chi, \sigma)$ 
space and take the coordinate with a negative eigenvalue 
as ``time". The metric \cb\ is in fact already in diagonal form. Near the 
event horizon \ch\ , 
\eqn\ci{
ds^2_5 = \Delta^{1/3}_{0+} ( {dx^2\over 4X}  + {XY_+\over U_+}d\sigma^2)+\ldots~,
}
where (we assume) $U<0$ at the horizon. 
Taking the time and the azimuthal angles imaginary, there is a conical singularity in \ci\ at the event horizon $X=0$ unless $\sigma$ has the
imaginary period 
\eqn\cj{
\beta_\sigma =  \left. {2\pi\over \partial_x X} \sqrt{\left|{U\over Y}\right|}\right|_{x=x_+}~.
}
The regularity condition was determined while keeping the space transverse to $\sigma$
fixed. Thus $\tilde{\chi}=\chi - \left.{Z\over U}\right|_{x=x_+}\sigma$ was fixed,
implying the imaginary period
\eqn\cka{
\beta_\chi  =\left. {Z\over U} \right|_{x=x_+}\beta_\sigma~,
}
and $\tilde{t} = t + {\cal A}_\chi \chi + {\cal A}_\sigma \sigma$ was fixed, 
giving the inverse Hawking temperature
\eqn\ck{
\beta_H  = \left.- {{\cal A}_\chi Z + {\cal A}_\sigma U\over U}\right|_{x=x_+}\beta_\sigma
=\left. {2\pi\over \partial_x X} {{\cal A}_\chi Z + {\cal A}_\sigma U\over \sqrt{-UY}}\right|_{x=x_+}~.
}

The angular velocity of the black hole along the angles $\sigma$, $\chi$
are simply the ratios
$\Omega_\sigma = \beta_\sigma/\beta_H$ and
$\Omega_\chi = \beta_\chi/\beta_H$. The rotational velocities along 
$\phi \pm \psi$ are then obtained from the linear transformations given in 
\ce. They are
\eqn\cl{\eqalign{
\Omega_R & = (\Omega_\sigma - \Omega_\chi)(b+a)
 = {(\beta_\sigma - \beta_\chi)\over \beta_H} (b+a)= 
 \left. {U-Z\over {\cal A}_\chi Z + {\cal A}_\sigma U} \right|_{x=x_+}(b+a)~,\cr
 \Omega_L & = (\Omega_\sigma + \Omega_\chi)(b-a)
 = {(\beta_\sigma + \beta_\chi)\over \beta_H} (b-a)= \left. {U+Z\over {\cal A}_\chi Z + {\cal A}_\sigma U} \right|_{x=x_+}(b-a)~.
}}

We also need the black hole entropy, extracted from the 
area of the event horizon:
\eqn\area{A_+=
\int dy\,   d\sigma\, d\chi \, \sqrt{{\rm det}\, g_{{\hat i}{\hat j}}}~,
}   
where $g_{{\hat i}{\hat j}}$ are the metric components 
of the $\{{\hat i}, {\hat j}\}=\{y,\sigma,\chi\}$ coordinates, evaluated at the 
outer  horizon $x=x_+$. The sub-determinant simplifies significantly
for any metric with the structure \cb: 
 \eqn\det{
 {\rm det}\, g_{{\hat i}{\hat j}}=g_{yy} {\rm det} (g_{\sigma\sigma}g_{\chi \chi}-g_{\sigma \chi}^2) =-{1\over {4YU}} \left({\cal A_\chi}Z+{\cal A_\sigma}U\right)^2~.
 }
Importantly it is {\it independent} of the conformal factor $\Delta_0$ 
(as well as $G$). 
 
In summary, in this subsection we have analyzed a general geometry of 
the form \cb\ without specifying all the functions \cd\ in detail. We found 
the position of the ergosphere \cf, the position of the event horizon \ch, 
the Hawking temperature \ck, the rotational velocities \cl,
and the black hole entropy \area. The main lesson from this computation is the remark that all these physical properties are independent of the warp factor $\Delta_0$. Accordingly we interpret the warp factor as a property of the surrounding spacetime, and not of the black hole ``itself". 

\subsec{Explicit Expressions}
The geometry we are interested in is specified by the functions \cd. In this case
the horizons at the roots of \ch\ are conveniently expressed
as
\eqn\ct{
x_\pm = {1\over 4} [ \sqrt{ \mu - (a-b)^2} \pm \sqrt{\mu - (a+b)^2}]^2~.
}
We mention in passing that the formulae in the preceding subsection all focussed on the event horizon $x=x_+$. Similar formulae clearly 
apply at the Cauchy horizon $x=x_-$. We have stressed the significance
of the inner horizon in earlier work 
(including \refs{\CveticUW,\LarsenGE}) and will not comment
further on this point in the present work. 

For the explicit geometries specified by \cd\ 
the inverse Hawking temperature \ck\ becomes
\eqn\cta{
\beta_H =  {1\over 2} ( \beta_R + \beta_L)~,
}
where
\eqn\cx{\eqalign{
\beta_R & = {2\pi\mu\over\sqrt{\mu-(b+a)^2}}\left( \Pi_c + \Pi_s\right)~,\cr
\beta_L & = {2\pi\mu\over\sqrt{\mu-(b-a)^2}}\left(\Pi_c - \Pi_s \right)~.
}}
The combination
\eqn\ctf{
{\cal A}_\chi Z + {\cal A}_\sigma U = \mu Y (\Pi_c x + ab \Pi_s)~,
}
is useful in the computations. 

The rotational velocities \cl\ become
\eqn\cxa{\eqalign{
\beta_H\Omega_R &= {2\pi (b+a)\over\sqrt{\mu-(b+a)^2}}~,\cr
\beta_H\Omega_L &= {2\pi (b-a)\over\sqrt{\mu-(b-a)^2}}~.
}}

The general formulae for the potentials \ck, \cl\ apply at any value of the polar 
angle $y$. The thermodynamic potentials should not depend on $y$
and our final expressions \cta, \cxa\ indeed do not. This constitutes a check on our computations.  

The black hole entropy computed from \area\ becomes 
$S_{\rm BH}={A_+\over{4G_5}}= S_R+S_L$ where
(in $G_5={\pi\over 4}$ units):
\eqn\gd{\eqalign{
S_L & 
= \pi\mu \sqrt{\mu - (b-a)^2}  \left( \Pi_c + \Pi_s\right)= 2\pi \sqrt{
{1\over 4} \mu^3 \left( \Pi_c + \Pi_s\right)^2-J^2_L}~,\cr
S_R & = 
\pi\mu \sqrt{\mu - (b+a)^2}  \left( \Pi_c- \Pi_s\right)= 2\pi \sqrt{{1\over 4} \mu^3\left( \Pi_c - \Pi_s\right)^2-J^2_R}~.
}}
The $y$ dependence of the subdeterminant \det\ cancelled entirely
prior to the integration \area. 

\newsec{The Near Horizon Geometry}
For near extreme black holes, the AdS/CFT correspondence can be derived
as a limit that decouples excitations in the near horizon region from the asymptotic geometry. This standard construction does not generalize to 
black holes that are not near extremality. This obstruction is physical: 
generally modes localized near the horizon couple to those 
in the asymptotic region. 

In this section we propose an alternative construction that addresses
this obstacle for general black holes. 

\subsec{The Subtracted Geometry}
The mechanics of our proposal is to modify the warp factor in the 
geometry \cb\ as
\eqn\da{
\Delta_0 \to \Delta~,
}
while maintaining all other aspects of the geometry. We will refer to the resulting metric as the {\it subtracted} geometry. 

It was shown in section 2 that thermodynamic potentials are independent 
of the warp factor. We interpret this to mean that the substitution \da\ 
leaves the {\it interior of the black hole unchanged}. The feature that 
changes is the asymptotic behavior of the geometry far from the 
black hole. This reflects a change the {\it environment} of the black hole.
We will choose the specific $\Delta$ in the subtracted geometry such that couplings between the black hole and modes far away are suppressed. 

In the context of the explicit solution \cd\ the subtraction we propose 
modifies the warp factor from
\eqn\dba{
\Delta_0 = \prod_{i=1}^3 (x+y + \mu\sinh^2\delta_i)~,
}
to
\eqn\db{
\Delta = \mu^2 \left[
(x+y) (\Pi^2_c -\Pi_s^2) +\mu\Pi_s^2 \right]~.
}
We will motivate this choice below, by imposing boundary conditions 
and the requirement that the wave equation remains separable in the subtracted geometry . 

The 5D geometries \cb\ asymptote flat space for large $x=r^2$. 
We verify this by estimating $\Delta_0 \sim x^3=r^6$, $G\sim r^4$, ${\cal A}\sim 0$, and $ds^2_4\sim r^{-2}\RR^4$. The subtracted warp factor \db\ increases
less fast, as $\Delta\sim x$. 
Consequently, the red-shift $g_{tt}\sim r^{8/3}$ of the 
subtracted geometry rises rapidly for large $x$. This prevents particles 
with finite energy near the black hole from escaping to infinity. We 
therefore interpret the subtracted geometry as the near horizon geometry of 
the black hole.

The subtracted geometry will generally not satisfy the equations of motion 
unless we also modify the matter supporting the original geometry. 
For example, the uncharged rotating solution (the Myers-Perry black hole)
no longer satisfies the vacuum Einstein equations after the warp factor
is changed from \dba\ to \db. In this situation the Einstein 
equation acting on the subtracted geometry determines the matter needed to support
the solution. We interpret such additional matter as the physical matter supporting the ``wall" that
we have introduced to separate the interior of the black hole from 
the irrelevant modes far away. 

Although we will not need to specify explicitly what matter is needed to support the solution we briefly pursue one approach that determines it, in section 3.4.

\subsec{The Wave Equation}
It is instructive to probe the geometry by a spectator scalar field satisfying the 
Klein-Gordon wave equation
\eqn\dkg{
[{1\over\sqrt{-g_5}} \partial_\mu (\sqrt{-g_5} g_5^{\mu\nu} \partial_\nu)  - M^2]\Phi=0~.
}
The inversion of the metric needed to render this differential equation explicit 
for the solution we consider is quite nontrivial; so we defer the details to 
an appendix. When the dust has settled, probes of the form 
\eqn\dc{
\Phi \sim e^{-i\omega t+ i m_R(\phi+\psi)
+im_L (\phi-\psi)}~,
}
are found to satisfy the equation 
\eqn\dd{\eqalign{&
\left[ 4\partial_xX\partial_x
+ 
{x_+-x_-\over x- x_+}\left( {\beta_R\omega\over 4\pi}  
- m_R{\beta_H\Omega_R\over 2\pi} + {\beta_L\omega\over 4\pi} - 
m_L{\beta_H\Omega_L\over 2\pi }\right)^2 \right. \cr
& - {x_+-x_-\over x- x_-}\left( {\beta_R\omega\over 4\pi } 
- m_R{\beta_H\Omega_R\over 2\pi} - {\beta_L\omega\over 4\pi }  +
m_L{\beta_H\Omega_L\over 2\pi}\right)^2   + \mu\omega^2 ( 1 + \sum_i\sinh^2\delta_i)
+x\omega^2 \cr
&+ \left. 4 \partial_y Y \partial_y + y\omega^2 + {1\over Y}\left( (a^2 + b^2 - y) \partial^2_\chi
+ y\partial^2_\sigma + 2ab\partial_\sigma \partial_\chi \right) 
+ {\Delta-\Delta_0\over G} \omega^2 \right]\Phi=M^2 \Delta^{1/3} \Phi~.
}}
The thermodynamic potentials $\beta_{L,R,H}, \Omega_{R,L}$ were 
given in \cx. The wave equation \dd\ pertains to arbitrary warp factor $\Delta$ but
we wrote the last term on the LHS with explicit reference to the unsubtracted
warp factor $\Delta_0$ so that the asymptotically flat 
black hole is easy to recover. 

{\it Separability} requires that the terms group into some that depend on the radial coordinate $x$ only and some that depend on the polar angle $y$ only. The first two lines in \dd\ are compatible with separability, as are most
terms in the last line of \dd. In fact, taking $M^2=0$ (for now), the only term that obstructs 
separability is the last term on the LHS of \dd. 
Separability is a striking characteristic of many rotating black hole solutions so 
we will elevate it to a principle that determines the possible warp factors. 

The massless wave equation \dd\ will be separable for any warp factor 
$\Delta$ such that 
\eqn\de{
{\Delta-\Delta_0\over G}  =  f_1(x) + f_2(y)  ~.
}
The functions $f_{1,2}$ are arbitrary at this point but they are constrained by boundary conditions: 
\item 
~i) $f_1(x)$ should have no poles at finite radius, or else the subtraction procedure will have introduced additional horizons. 
\item
~ii) $f_1(x)$ rising faster than linear for large $x$ {\it increases} 
couplings between the black hole region and asymptotic space.
\item
~iii) The linear coefficient in $f_1(x)$ just changes normalization of the asymptotic Minkowski space, except in the special case of slope ``-1'' where
it suppresses couplings.
\item
~iv) Constants in $f_{1,2}$ can be absorbed into separation constants. 
 
\noindent
These considerations motivate taking $f_1(x)= -x + {\rm const.} $, since this is
the only case that will make escape to infinity more difficult, while 
preserving separability and analyticity. Taking $f_2(y)$ linear as well 
(so that $\Delta$ is a function of the combination $x+y$)
and choosing separation constants conveniently we are lead to take
\eqn\deg{
\Delta =  \Delta_0  - [ x + y + \mu ( 1 + \sum_i\sinh^2\delta_i) ] G ~.
}
This is precisely the warp-factor $\Delta$ announced in \db. It is interesting that the warp factor remains unchanged at the static limit \cf\ separating the ergosphere from the asymptotic space. 

The terminology referring to a ``subtracted'' geometry is motivated by 
\deg: the warp factor $\Delta$ of the subtracted geometry is expressed as the 
warp factor $\Delta_0$ of the asymptotically flat black hole, less terms that
are separable. 

With our choice of warp factor the wave equation \dd\ in the subtracted 
geometry separates into the angular Laplacian on the round three 
sphere $S^3$:
\eqn\dy{
\left[ 4 \partial_y Y \partial_y  + {1\over Y}\left( (a^2 + b^2 - y) \partial^2_\chi
+ y\partial^2_\sigma + 2ab\partial_\sigma \partial_\chi \right) \right] \Phi_\Omega= - j(j+2) \Phi_\Omega~,
}
and a radial equation
\eqn\dx{\eqalign{&
\left[ 4\partial_xX\partial_x
+ 
{x_+-x_-\over x- x_+}\left( {\beta_R\omega\over 4\pi}  
- m_R{\beta_H\Omega_R\over 2\pi} + {\beta_L\omega\over 4\pi} - 
m_L{\beta_H\Omega_L\over 2\pi }\right)^2 \right. \cr
& \left.  - {x_+-x_-\over x- x_-}\left( {\beta_R\omega\over 4\pi } 
- m_R{\beta_H\Omega_R\over 2\pi} - {\beta_L\omega\over 4\pi }  +
m_L{\beta_H\Omega_L\over 2\pi}\right)^2 \right]\Phi_r=j(j+2) \Phi_r~.
}}
The separation constant $j$ is just the usual angular momentum quantum number. The azimuthal quantum numbers are bounded as
$|2m_R|, |2m_L| \leq j$. 

The radial wave equation \dx\ for the subtracted geometry is just the hypergeometric equation, with singular points at the outer and inner
black hole horizons and at asymptotic infinity. The well-known relation 
between the hypergeometric equation and $SL(2,\RR)$ is the first step 
towards identifying a Virasoro algebra in the general subtracted geometry. 

\subsec{Approximation vs. Subtraction}
The simplified radial equation \dx\ is usually interpreted as an 
{\it approximation} to the full answer, applicable when the linear terms 
in \dd\ are negligible compared with the remaining terms. This 
interpretation can be justified  in many situations that involve a small parameter, such as: 
\item 
~i)
The dilute gas regime (hierarchy between the charges) 
\refs{\MaldacenaIX,\CveticUW}.
\item 
~ii)
The near-extremal Kerr regime (large angular momentum) \refs{\BredbergPV,\CveticXH}.
\item
~iii)
Small probe energy compared to all black hole parameters 
\refs{\CveticUW,\CastroFD}.

\noindent
Our procedure changes the geometry from the outset, by introducing the subtracted warp factor \db, and then computes the exact wave equation. 
Although the result is the same, our approach brings several advantages:
\item 
~i)
The estimates justifying application of the simplified radial 
equation \dx\ generally require assumptions about the black hole 
parameters, such as near extremality of the black hole. 
Those assumptions we sidestep here, and so our result applies to the 
general family of black holes.  
\item 
~ii).
The geometry corresponding to the simplified radial equation is exhibited explicitly, rather than on the level of the wave equation. This facilitates a more 
thorough analysis, such as generalization to probes with spin. 

\subsec{Moduli Space}
The subtracted geometry does not generally satisfy the equations of motion. Our attitude to this situation is that any enclosure of the black hole necessarily must be formed from matter, and the equations of motion then specify what kind of matter is needed. 

Despite this philosophy it is instructive to identify suitable matter
by judicious exploitation of moduli space. To do so we start with the black hole
at a general point in moduli space, where the warp-factor is
\eqn\fk{
\Delta_0 = (x+y)^3 \prod_{i=1}^3 ( h_i + {\mu \sinh^2 \delta_i\over x+y})~.
}
In this case the geometry is a solution with the usual $N=2$ matter specified in terms of harmonic functions ${\cal H}_i$ with constant terms $h_i$. 

At this point we have a family of geometries with specified matter and 
warp factors parametrized by the values of $h_i$. The next step is to 
adjust the moduli such that ($s_i \equiv \sinh\delta_i$)
\eqn\ea{\eqalign{
h_1 h_2 h_3 & = 0~,\cr 
 h_1 h_2 s_3^2 + h_2 h_3 s_1^2   + h_3 h_1 s_2^2  & = 0~,\cr
h_1 s^2_2 s^2_3 + h_2 s^2_1 s^2_3 +h_3 s^2_1 s^2_2   & 
=\Pi_c^2 - \Pi^2_s  ~.
}}
At this point in moduli space the warp factor \fk\ takes the ``subtracted'' form \db. 

For given boost parameters $\delta_i$, the conditions \ea\ constitute three equations for three variables so 
we expect that they have a solution. Indeed, we can find solutions by breaking the cyclic symmetry between 
the three charges. One of the solutions is
\eqn\eb{\eqalign{
h_1 & = h_2 =0~, \cr
h_3 & =  {1\over s_1^2 s_2^2} (\Pi_c^2 - \Pi^2_s)~. 
}}
This is evidently a rather singular point in moduli space, as one would expect since it corresponds to a change in asymptotic behavior. Indeed, it can be interpreted as a limit where two physical charges are taken large with the third finite; and this is just the standard decoupling limit that identifies an AdS$_3$ near horizon geometry in the 5D black hole.

As we have stated repeatedly, the precise matter supporting the boundary conditions are in fact not important to us. The virtue of the implicit construction of suitable matter, by taking a singular limit in moduli space, is that it provides evidence that the required matter will in fact physically sensible. It also guarantees that our general computation will reduce to standard AdS/CFT
results (such as \CveticXH) when two charges are large.

\newsec{Linear Realization of Conformal Symmetry}
Let us summarize the situation up to this point: the wave equation in the general black hole geometry \cb\ has several nice properties, such as separability; but it does not quite reduce to hypergeometric form. This is remedied in the ``subtracted'' geometry, where the warp factor has been changed from \dba\ to \db. 

Now, the hypergeometric wave equation ensures that there are 
$SL(2,\RR)^2$ generators acting on scalar probes of the subtracted geometry. 
We want to understand the nature of this symmetry and, if possible, extend it to a full conformal symmetry. An illuminating way to proceed is to embed the subtracted 5D black hole geometry in an auxiliary 6D geometry. 

\subsec{The Auxiliary 6D Geometry}
The subtracted geometry remains intricate, as one expects for a 
charged rotating black hole. The main complication is that the radial 
coordinate $x$ and the polar coordinate $y$ couple extensively. 
In appendix A we recast the full metric in a form with radial/temporal 
terms that depend on $x$ only, angular terms that depend on $y$
only, and a single term encoding the coupling between $x$ and $y$. 
In appendix B we show how the remaining coupling between $x$ and $y$
can be eliminated as well, by introducing an auxiliary direction 
parametrized by a coordinate $\alpha$. The 6D auxiliary
geometry resulting from this construction is 
\eqn\fa{\eqalign{
\ell^{-2} ds^2_6 & = \Delta ({1\over\mu}d\alpha + {\cal B})^2 + \Delta^{-1/3}ds^2_5 \cr
& =  \Delta ({1\over\mu}d\alpha + {\cal B})^2 - \Delta^{-1} G ( dt+{\cal A})^2 + ds^2_4~,
}}
where the KK-field along $\alpha$ is
\eqn\faa{\eqalign{
{\cal B}& =  {1\over \Delta} \left[  \mu ((a^2 + b^2-y) \Pi_s -ab\Pi_c )d\sigma + 
\mu (y\Pi_c-ab\Pi_s )d\chi -  {\Pi_s\Pi_c \over \Pi^2_c-\Pi_s^2}
dt  \right]~.
}}
The arbitrary 
length scale $\ell^{-2}$ was introduced on the LHS of \fa\ in order that 
dimensions work out correctly. 

The auxiliary direction is just a formal device (for now): the 5D wave 
functions are in 1-1 correspondence with 6D wave functions that are
independent of $\alpha$. To see this note that
\eqn\fb{
\ell^{-4}\sqrt{-g_6} g^{\mu\nu}_6 = \sqrt{-g_5} g^{\mu\nu}_5~,
}
for $\mu, \nu$ in the 5D space. Massless 6D fields independent of $\alpha$ thus satisfy precisely the same wave equation as massless 5D fields. 

The 6D representation \fa\ is aesthetically pleasing in that the awkward fractional powers of the warp factor $\Delta$ have been removed. A related
physical simplification is the trivial overall conformal factor $g_6=-\ell^{12}$. 
It implies that {\it massive} fields coupling minimally to
the 6D metric
\eqn\fcf{
[{1\over\sqrt{-g_6}} \partial_I (\sqrt{-g_6} g_6^{IJ} \partial_J)  - 
M^2_6]\Phi=0~,
}
satisfy a separable wave equation of hypergeometric form. This 
contrasts with massive fields coupling minimally to the 5D metric: 
the RHS of \dd\ obstructs separation of variables by coupling radial and 
angular directions.\foot{Massive particles in the unsubtracted geometry with 
{\it diagonal} charges separate in 5D as well since then $\Delta_0$ is 
the cube of a linear function; but this case does not reduce to 
hypergeometric form \ChongZX.} 

In section 2 we determined the physical temperature of the black hole using a regularity condition at the Euclidean horizon. The computation applies 
to the 6D geometry \fa\ as well. The condition that the line element 
$\alpha+\mu {\cal B}$ be kept fixed as the horizon is circumnavigated
then determines the periodicity
of $\alpha$ as
\eqn\fza{\eqalign{
\beta_\alpha &= -\mu ({\cal B}_\sigma \beta_\sigma+
{\cal B}_\chi \beta_\chi + {\cal B}_t \beta_H)_{x=x_+}\cr
&= {\Pi_s + {ab\over x_+} \Pi_c\over \Pi_c^2 - \Pi^2_s}\beta_\sigma\cr
&= {\pi\over (\Pi_c - \Pi_s)\sqrt{\mu - (b+a)^2}}- 
{\pi\over (\Pi_c + \Pi_s)\sqrt{\mu - (b-a)^2}}~.
}}
The expression \faa\ for ${\cal B}$ depends nontrivially 
on the polar coordinate 
$y$ but such dependence cancels in \fza, as it should for the 
thermodynamic potential $\beta_\alpha$. 
This gives a non-trivial check on our computations. 

\subsec{Factorization}
As advertized in the beginning of this section, the 6D 
geometry \fa\ separates variables 
manifestly. To see this, we simply expand the functions and collect 
terms (some details are in the appendices). We find
\eqn\fxd{\eqalign{
\ell^{-2} ds^2_6 &=   - {X\over \mu^2 S} dt^2  + {dx^2\over 4X} + S(d\alpha - 
{q_t\over S} dt)^2  + 
{dy^2\over 4Y}+ {Y\over y}d\tilde{\sigma}^2 + y ( d\tilde{\chi} - {ab\over y}d\tilde{\sigma})^2~,
}}
where $S$ is a linearly transformed version of the radial coordinate $x$ 
\eqn\fxda{
S =  x (\Pi_c^2 - \Pi^2_s)  + 2ab \Pi_c \Pi_s -(a^2 + b^2 - \mu)\Pi^2_s~,  
}
the potential $q_t$ is
\eqn\fxdb{
q_t = - {ab (\Pi_c^2 + \Pi^2_s) - (a^2 + b^2 - \mu)\Pi_s\Pi_c\over
\mu (\Pi_c^2 - \Pi^2_s)}~,
}
and the shifted azimuthal coordinates are,
\eqn\fxe{\eqalign{
d\tilde{\chi} &= d\chi - {\Pi_s\over \mu (\Pi^2_c - \Pi^2_s)}dt  + \Pi_c d\alpha~,\cr
d\tilde{\sigma} &= d\sigma - {\Pi_c\over \mu (\Pi^2_c - \Pi^2_s)}dt + \Pi_s d\alpha~.
}}
Not only does \fxd\ separate variables manifestly directly in the geometry:
the metric is locally AdS$_3\times S^3$ even for the general black holes we study. 
This is important because it immediately implies that the $SL(2,\RR)^2$ isometries are enhanced to Virasoro algebras
\refs{\BrownNW,\StromingerEQ}. This is what we wanted to show. 

The generators of the AdS$_3$
Virasoro algebras
depend in an essential manner on the spatial isometry parametrized by 
the coordinate $\alpha$. In physical terms, the oscillations that the Virasoro algebras act on are right and left moving waves moving along the $\alpha$ directions. The status of such states is uncertain since $\alpha$ was introduced as an auxiliary variable. 

Having mentioned this important caveat, we should also recall that once 
there is an AdS$_3$ component of the geometry, the entropy is in fact 
{\it guaranteevd} to work out: modular invariance in the CFT is geometrized directly in the BTZ black hole \refs{\DijkgraafFQ,\KrausVZ}, and so the black hole entropy is computed
by the low energy thermal modes in a manner that always gives the ``right'' result. The important point will therefore not be to establish a numerical agreement for the black hole microstates, but rather to understand whether these states are in fact physical. 

\subsec{BTZ Interpretation}
It is instructive to express the first three terms in \fxd\ in the standard BTZ
form
\eqn\fya{
ds^2_{\rm BTZ} = - N^2 dt^2 + N^{-2} dR^2 + R^2 ( d\phi_{\rm BTZ} + 
{4G_3J_3\over R}dt)^2~,
}
where
\eqn\fyb{
N^2 = {(R^2 - R^2_+)(R^2 - R^2_-)\over \ell^2 R^2}= 
{R^2\over \ell^2} - 8G_3M_3 + {16G^2_3 J^2_3\over R^2}
~.
}
Comparison with \fxd\ identifies the time coordinates and gives simple
rescalings for the remaining coordinates 
\eqn\fyc{\eqalign{
S &= {R^2 \mu^2 (\Pi^2_c - \Pi^2_s)^2\over\ell^4}~,\cr
\phi_{\rm BTZ} & = {\mu (\Pi^2_c - \Pi^2_s)\over\ell} \alpha  ~.
}}
The horizon loci $R^2_\pm$
%
%
give the effective BTZ black hole parameters
\eqn\fye{\eqalign{
8G_3 M_3 &= {\ell^2\over \mu^2  (\Pi^2_c - \Pi^2_s)^2}
\left[  ( \mu - a^2-b^2) (\Pi^2_c + \Pi^2_s)+
4ab\Pi_c\Pi_s\right]~,\cr
4G_3 J_3 &= {\ell^3\over
\mu^2 (\Pi_c^2 - \Pi^2_s)^2}
[ab (\Pi_c^2 + \Pi^2_s) - (a^2 + b^2 - \mu)\Pi_s\Pi_c]~.
}}

\newsec{Black Hole Microstate Counting}
The $SL(2,\RR)^2$ symmetry of the subtracted geometry is 
non-abelian so all generators are normalized uniquely. The hidden conformal symmetry approach combines this property with the known periodicity of the azimuthal angles and infer notions of temperature in the dual CFT 
description. 
Alternatively, we can take advantage of the auxiliary coordinate we have introduced to identify an explicit Virasoro algebra using standard AdS/CFT technology. 
In this section we develop both these approaches. 

\subsec{Hidden Conformal Symmetry}
The non-abelian nature of $SL(2,\RR)^2$ determines the properly 
normalized $U(1)$ generators as\foot{The properly normalized 
generators were introduced already in \CveticUW. Eqs. 56 and 57 
of \CveticUW\ are ${\cal R}_3, {\cal L}_3$, and the remaining generators 
are given in the preceding equations.}
\eqn\ga{\eqalign{
{\cal R}_3 & = {i\over 4\pi}\left( \beta_R \partial_t + \beta_H \Omega_R (\partial_\phi+\partial_\psi)\right)~,\cr
{\cal L}_3 & = {i\over 4\pi} \left(\beta_L \partial_t + \beta_H \Omega_L (\partial_\phi -\partial_\psi)\right)~.  
}}
We may realize these operators as 
${\cal R}_3=i\partial_{t_R},~{\cal L}_3=i\partial_{t_L}$ 
by introducing the coordinates
\eqn\gb{\eqalign{
t_R & = {4\pi\over\beta_R}\left( t - {\beta_L\over 2\beta_H\Omega_L}( \phi - \psi)\right)~,  \cr
t_L & = {4\pi\over\beta_L}\left(t - {\beta_R\over 2\beta_H\Omega_R}( \phi + \psi)\right)~. 
}}
The ``hidden" part of hidden conformal symmetry refers to the observation that such coordinates are globally ill-defined, because of periodicity along the azimuthal angles. Comparing the periodicities of $t_R, t_L$ obtained this way with the standard CFT definition of temperature 
$z\equiv z + 4\pi^2 i T^{\rm CFT}$ we find
\eqn\gc{\eqalign{
T^{\rm CFT}_R & = {1\over\beta_R} {\beta_L\over\beta_H\Omega_L}~,\cr
T^{\rm CFT}_L & = {1\over\beta_L} {\beta_R\over\beta_H\Omega_R}~.
}}
Note that these dimensionless CFT temperatures differ from the 
dimensionfull physical 
temperatures $T_{R,L}^{\rm phys}=\beta_{R,L}^{-1}$ that govern 
Hawking radiation. 

If a CFT with the temperatures \gc\ is responsible for the black hole 
entropy \gd\ it must be that 
\eqn\ge{\eqalign{
S_R &= {\pi^2\over 3}c_R T^{\rm CFT}_R =  
\pi\mu\sqrt{\mu-(b+a)^2}
\left( \Pi_c- \Pi_s\right)~.
}}
Simplifying this expression using the parametric expressions for 
potentials \cx,\cxa\ and the physical black hole parameters \cg\ we find the central charge
\eqn\gea{
c_R = 12 |J_L|~.
}
Similarly, $S_L$ works out correctly if $c_L=12|J_R|$. 

The apparent interchange up of $R$ and $L$ is surprising but it could be correct. For example, in the extremal limit where $S_L\to 0$ with no charges
the full entropy
$S_R = 2\pi\sqrt{J^2_L - J^2_R}$ which is well described by a CFT with $c_R=12|J_L|$ and R-charge identified with $J_R$. Likewise it is also acceptable {\it a priori} that the right and central charges are different such 
that the underlying theory is chiral. This could well be a feature required for a CFT describing black holes with angular momenta $J_R\neq J_L$ and arbitrary charges. 

An important challenge to the hidden conformal symmetry program is 
the non-uniqueness of the coordinates \gb\ realizing the conformal generators. The specific realization \gb\ is such that symmetry between R and L is preserved but it would be equally acceptable to take
\eqn\geb{
t'_L = {2\pi\over\beta_H\Omega_L}( \phi - \psi)~,
}
while keeping the $t_R$ in \gb. In this basis the temperatures are 
\eqn\gec{
T'^{\rm CFT}_L = {1\over\beta_H\Omega_L}~,
}
with $T^{\rm CFT}_R$ still given by \gc. Assuming again that the entropies
\gd\ are reproduced in the CFT, the central charges are inferred as $c_R=c_L=12|J_L|$. This basis thus suggests a non-chiral CFT.

It is clear that many bases realize the conformal generators and that their periodicity conditions motivate a range of central charges. The primed basis above is natural in the near extreme limit of a rotating black hole with all rotation along the $J_L$ direction but it is clearly an awkward choice for 
black holes rotating mostly along the $J_R$ direction. This ambiguity 
challenges the proposal that a CFT with these central charge assignments
might account for the entropy of all black holes. 

\subsec{The Long String Picture}
Our embedding of the black hole geometry into 6D suggests another avenue for understanding the black hole entropy. In this approach the excitations responsible for the entropy are along the auxiliary coordinate $\alpha$ 
rather than the azimuthal angles in physical space. To develop this scenario 
we must specify the AdS$_3$ radius $\ell$ which is not determined 
by the wave equation and also compute the effective 3D coupling 
$G_3$ which we have kept arbitrary for now. 

We first rewrite the answer we seek as follows. We describe R and L 
entropies \gd\ in terms of dilute gasses with physical 
temperatures \cx\ so 
\eqn\fyga{
S_{R,L} = {\pi^2\over 3} c T^{\rm phys}_{R,L}{\cal R}~,
}
where for both R and L chiralities
\eqn\fygb{
c {\cal R} = 6\mu^2 ( \Pi^2_c - \Pi^2_s)~.
}
The length scale $2\pi {\cal R}$ is the volume of the 1D gas, needed to transform from physical temperatures to the dimensionless CFT temperature. A microscopic understanding amounts to accounting for \fygb. 

The effective 3D coupling constant $G_3$ is identified by comparing 
the 6D$\to$ 3D reduction on $S^3$ with the 6D$\to$ 5D reduction on the
circle parametrized by $\alpha$ \BalasubramanianEE. Assigning again 
the length scale $2\pi {\cal R}$ to the effective string direction we 
find
\eqn\fyf{
G_3 = G_5 {2\pi {\cal R}\over V(S^3)} = {4G_5\over\pi}\cdot{{\cal R}\over 4\ell^3}~.
}
The Brown-Henneaux formula then gives
\eqn\fyff{
c {\cal R} = {3\ell\over 2G_3} \cdot {\cal R} = 6\ell^4~,
}
in string theory conventions (see \LarsenXM) where $G_5 = {\pi\over 4}$. 
Combining formulae, \fygb\ implies the AdS$_3$ scale
\eqn\fyaa{
\ell^4 = \mu^2 ( \Pi_c^2 - \Pi_s^2)~.
}
The length scale ${\cal R}$ drops out, as it should due to conformal invariance in the boundary theory. The invariant scale that we need to explain is 
the AdS$_3$ radius \fyaa.

The key ingredient we have at our disposal is the result from section 2.2 
that the thermodynamic potentials and the black hole entropy are independent of the warp factor and so on the subtraction procedure. 
Since we have assigned periodicity $2\pi{\cal R}$ to the periodic length scale 
in the CFT, the BTZ angle $\phi_{\rm BTZ}$ has periodicity 
$2\pi {\cal R}/\ell$; and so the BTZ entropy becomes
\eqn\fyg{\eqalign{
S_{\rm BTZ} &= {A_3\over 4G_3} = 
{2\pi R_+ \over 4G_3}\cdot {{\cal R}\over\ell} \cr
&= 2\pi\cdot {\ell^4\over\mu(\Pi^2_c-\Pi_s^2)}\cdot
\left( \sqrt{\mu - (a-b)^2}(\Pi_c + \Pi_s)+
\sqrt{\mu - (a+b)^2}(\Pi_c - \Pi_s)\right)~.
}}
Consistency with the 5D black hole entropy \gd\ then implies \fyaa. This 
inference constitutes a {\it derivation} of the AdS$_3$ length scale which in turn accounts for the black hole entropy in full generality. 

At this point we can apply standard AdS/CFT technology to derive further results. For example, the conformal weights assigned to the black hole are 
$h_{L,R} = {M_3\ell \pm J_3\over 2}$ where the BTZ parameters are \fye.

A particularly appealing feature of the AdS$_3$ description is the manifest modular invariance, realized geometrically as interchange of the (Euclidean) 
$t$-direction and the auxiliary $\alpha$-direction \DijkgraafFQ. 
The symmetry between these directions is apparent already in the thermodynamic potential 
\fza\ which we can write as
\eqn\fyj{
\beta_\alpha = {1\over 2\mu(\Pi^2_c-\Pi^2_s)} ( \beta_R - \beta_L)~.
}
The relation $\fyc$ to $\phi_{\rm BTZ}$ which has known 
periodicity $2\pi{\cal R}/\ell$ then determines
\eqn\fyk{
{\cal R} = {\beta_R-\beta_L\over 4\pi}~.
}
This quantity is interpreted physically as a chemical potential.

The $S^3$ is fibered over the AdS$_3$ base space, as encoded in the shifted angles \fxe. We interpret this as a generalization of the corresponding effect for near extreme black 
holes \refs{\CveticXH,\MaldacenaDR,\BalasubramanianRT,\LuninIZ}: 
the effective string in the $(t,\alpha)$ plane has been boosted by $\cosh\delta_0 = \Pi_c/\sqrt{ \Pi_c^2 - \Pi_s^2}$, and the size of the sphere is set by the scale $\ell$ in \fyaa\foot{The shifts \fxe\ generalize eqs 13 and 14 in \CveticXH. They reduce precisely to those in the limit where $\delta_{1,2}\gg 1$.}. These shifts encode the energy cost of carrying angular momentum. 

For near extreme black holes with large charges there are different limits where the description becomes effectively free. In the case of the extreme 
D1-D5 system the asymptotic behavior in one limit is
controlled by energy $h_{\rm eff}= Q_1 Q_5 p$, with each string contributing of 
order one to the level; but in another limit the energy  $h_{\rm eff}=p$, with each 
string contributing fractions of order $1/Q_1Q_5$\MaldacenaDS. In the 
near extreme limit the scale \fyaa\ increases rapidly with boost parameter and reduces precisely to the ``long'' string scale, the one that 
gives rise to maximal fractionation. The description in this subsection is thus 
a generalization of the long string picture. 

The hidden conformal symmetry approach interprets the symmetries
differently from the long string picture.\foot{A technical difference that obscures comparisons is that hidden conformal symmetry is developed in the 
dual modular frame where temperature is identified as imaginary periodicity $4\pi^2 T$, while in the long string picture we identify the temperature with
the imaginary periodicity $\beta$.} In the non-chiral version of hidden 
conformal symmetry the division of the entropy into R and L parts agree, 
as do the corresponding assignments of temperatures. The remaining
physical difference is then the length scale that relates the physical 
temperature to the CFT temperature. Those are genuinely different.

\newsec{Discussion}
We conclude with a short discussion of our main results and some
future directions that they open: 
\item
~i)
{\it Subtracted Warp Factor:} we modify the warp factor while maintaining 
all the remaining parts of the metric. It would interesting to generalize this construction to other contexts.

\item
~We accepted that the subtracted metric does not satisfy the equations of motion, arguing that this is a physical property of adding an enclosure that decouples the black hole from the asymptotically flat space. It would be nice to understand the implied supporting matter in more detail. Alternatively, in 
some cases one may prefer to argue that the subtracted warp factor is a good approximation 
in the important region. 

\item
~ii)
{\it The 6D Lift:} we lifted the subtracted geometry to one dimension 
higher, by adding an auxiliary coordinate $\alpha$. 
This simplified the otherwise very complicated geometry enormously. We anticipate that this type of lift to  
higher dimensions will be useful in many contexts. For example, it may shed light on the mysterious symmetries enjoyed by the 4D Kerr black hole. 
The auxiliary coordinate also appears to create novel symmetry transformations  that mix it with the physical coordinates.

\item
~iii)
{\it Black Hole Entropy:} our construction provides a fairly systematic 
way to identify a conformal symmetry for a very large class of black 
holes. It should be possible to exploit this result to get a convincing account of the general black hole entropy. We developed two approaches: hidden conformal symmetry (adapting \CastroFD\ to the present setting) and what 
we refer to as the long string picture (detailing our earlier work, including 
\refs{\CveticUW,\LarsenGE}). In the latter approach we were able to find a quantitative match for the entropy with no free parameters. 

\bigskip
\noindent {\bf Acknowledgments:} \medskip \noindent
We thank A. Castro for discussions. We thank CERN for hospitality when this 
work was initiated. FL also thanks UPenn and CERN (again) for hospitality
as the work was completed. MC is supported by the DoE 
Grant DOE-EY-76-02- 3071, the NSF RTG DMS Grant 0636606, the 
Fay R. and Eugene L. Langberg Endowed Chair, and the Slovenian 
Research Agency (ARRS). FL is also supported by the DoE.

\appendix{A}{Derivation of the Scalar Wave Equation}
We are putting great emphasis on the form of the scalar wave equation.
The analogous simplifications are less evident directly in the geometry \cb. 
In this appendix we present the steps we take to determine the scalar 
wave equation from the subtracted geometry. 

The poles in the metric (and in the wave equation) provide significant 
guidance for the explicit manipulations. For example, the base 
metric $ds^2_4$ in \cd\ appears to have a pole where $U=0$ but, 
upon expansion of the terms, contributions to this pole in fact 
cancel. We can recast the base metric as
\eqn\wa{
ds^2_4 = {1\over G}\left[ U d\chi^2 + \left( (X+Y) (a^2 + b^2)- U - \mu Y
\right) d\sigma^2 - 2ab (X+Y)d\sigma d\chi\right]~.
}
The apparent pole where $G=0$ similarly cancels in the full 5D metric. 
To see this, introduce the coordinates
\eqn\wc{\eqalign{
S &=  x (\Pi_c^2 - \Pi^2_s)  + 2ab \Pi_c \Pi_s -(a^2 + b^2 - \mu)\Pi^2_s~,  \cr
T &= y (\Pi^2_c - \Pi^2_s) - 2ab \Pi_c \Pi_s+ (a^2 + b^2)\Pi^2_s ~.
}}
The coordinates $S, T$ are just linear transformations on the radial and polar coordinates $x, y$
(to which they reduce in the absence of charges). They pick out the radial and polar
part of the conformal factor
\eqn\wd{
\Delta = \mu^2(S+T)  = (\Pi^2_c - \Pi^2_s) (x+y) + \mu\Pi^2_s~.
}
After a straightforward (and only moderately tedious) computation we find
\eqn\we{\eqalign{
\Delta^{- 1/3}ds^2_5 & =
  - {G\over \mu^2(S+T)} (dt+{\cal A})^2 + ds^2_4\cr
& =
 - {G\over \mu^2(S+T)} (dt^2+2{\cal A}dt)
+ [ y d\chi^2 + (a^2 + b^2 - y) d\sigma^2 - 2ab d\sigma d\chi+   {dy^2\over 4Y}]  \cr
&  + {dx^2\over 4X} - {1\over S+T} [ (\Pi_c y - ab \Pi_s)d\chi - (\Pi_c ab - \Pi_s (a^2 + b^2 - y)) d\sigma]^2 ~.
}}
The apparent pole at $G=0$ cancelled as claimed. The angular terms take a nice form in
this equation: the first
square bracket is just the round sphere $S^3$, and the second one
then represents a deformation due to rotation and charges. 

To find the wave equation we need to invert the metric and for this purpose \we\ is
not optimal. A good alternate form (which takes some effort to reach) is
\eqn\wf{\eqalign{
\Delta^{- 1/3} ds^2_5 & = {dx^2\over 4X}   - {X\over \mu^2 S} dt^2 + {dy^2\over 4Y}+ {Y\over T}
(\Pi_s d\chi - \Pi_c d\sigma+ {1\over\mu}dt )^2 + {ST\over S+T} \left( {p\over T} - {q\over S}\right)^2~,
}}
where
\eqn\wg{\eqalign{
p& =   ((a^2 + b^2-y) \Pi_s -ab\Pi_c )d\sigma + 
(y\Pi_c-ab\Pi_s )d\chi+ { ab(\Pi^2_c+\Pi^2_s) - (a^2 + b^2)\Pi_s\Pi_c \over\mu(\Pi^2_c-\Pi_s^2)}
dt~,\cr
q& =  - {ab(\Pi^2_c+\Pi^2_s) - (a^2 + b^2-\mu)\Pi_s\Pi_c  \over\mu(\Pi^2_c-\Pi_s^2)} dt~.
}}
This form of the metric is a sum of five complete squares so it is explicitly diagonalized.
The inverse metric is therefore just the inverse of each eigenvalue, written in the dual basis:
\eqn\wh{\eqalign{
\Delta^{1/3} g^{\mu\nu}\partial_\mu\partial_\nu
&= 4X \partial^2_x -{S\over X} \left( \mu \partial_t - {1\over\Pi_s}\partial_\chi + {1\over S}
(x\Pi_c +ab\Pi_s )(\partial_\sigma + {\Pi_c\over\Pi_s}\partial_\chi)
\right)^2 + 4Y \partial^2_y
\cr
& + {T\over Y}  [{1\over\Pi_s}\partial_\chi - {y\Pi_c- ab\Pi_s\over T} (\partial_\sigma + {\Pi_c\over\Pi_s}\partial_\chi)]^2+
({1\over S} +{1\over T})(\Pi_s \partial_\sigma + \Pi_c\partial_\chi)^2~. 
}}
The determination of the dual basis uses the identity
\eqn\wi{\eqalign{
{y\over T} - {x\over S} 
& = {\Pi_s\over \Pi^2_c-\Pi_s^2} \left( { 2 ab\Pi_c - (a^2 + b^2)\Pi_s\over T} 
+ { 2 ab\Pi_c - (a^2 + b^2-\mu)\Pi_s\over S}\right)~.
}}

The determinant of the metric is just ${\rm det} g_5 = - {1\over 16}\Delta^{2/3}$ so the Laplacian
operator becomes
\eqn\wj{\eqalign{
&\partial_\mu(\sqrt{-g_5} g^{\mu\nu}\partial_\nu)
= \partial_x (X \partial_x) -{S\over 4X} \left( \mu \partial_t + {\Pi_s (x+ a^2 + b^2 - \mu) - ab\Pi_c\over S}\partial_\chi + {x\Pi_c +ab\Pi_s \over S}
\partial_\sigma 
\right)^2 
\cr
&\!\!\!\!\!\! \!\!\!+  \partial_y (Y \partial_y) + {1\over 4YT}  [((a^2 + b^2 -y)\Pi_s- ab\Pi_c) \partial_\chi - (y\Pi_c- ab\Pi_s)\partial_\sigma ]^2+
{1\over 4}({1\over S} +{1\over T})(\Pi_s \partial_\sigma + \Pi_c\partial_\chi)^2\cr
& =  \partial_x (X \partial_x) - {1\over 4X}\left[ S\mu^2 \partial^2_t
+ 2\mu (\Pi_s (x+ a^2 + b^2 - \mu) - ab\Pi_c)\partial_\chi\partial_t  + 
2\mu(x\Pi_c +ab\Pi_s)\partial_\sigma \partial_t \right.\cr
& \left. + x\partial^2_\sigma - (x+ a^2 + b^2 - \mu) \partial^2_\chi - 2ab \partial_\chi \partial_\sigma\right]
+  \partial_y (Y \partial_y)  + {1\over 4Y} [ (a^2 + b^2 - y) \partial^2_\chi + y\partial^2_\sigma
  + 2ab\partial_\sigma\partial_\chi] ~. 
}}
In the first expression we rewrote the terms so that the limit of vanishing charge $\Pi_s\to 0$ is 
manifestly regular. In the second expression we collected term to make is manifest that
there are no poles in the wave equation at $S=0$ and $T$. 
\medskip

\noindent
The final expression \wj\ has several notable features:
\item 
~i)
{\it Separation of variables:} the radial and angular variables do not couple. 
\item 
~ ii)
{\it Singularity structure:} as we have emphasized, most presentations of the metric and/or 
the wave equation has a number of spurious singularities (eg. at $U=0$, $G=0$, $S=0$, $T=0$) 
but these are all absent in \wj. 
\item 
~iii)
{\it Only simple poles:} $X$ given in \cd\ is a quadratic function with two distinct
roots (for nonextremal black holes). Thus the poles at $X=0$ can be decomposed such that the only singularities are simple poles. 
\item 
~iv)
{\it Spherical symmetry:}  the angular Laplacian (the square bracket in the last
line) is the same as in flat space. Thus rotation of the black hole has not 
broken rotational symmetry in the wave equation of the subtracted metric.
\item 
~v)
{\it Locality:} the expression in the first square bracket is {\it linear} in $x$.
This is the feature that suppresses coupling of modes to the asymptotic space.
Linearity means the term can be decomposed as exactly two pole terms. 
The complicated dependence on parameters in these terms
is just due to the intricate thermodynamics of these black holes, as 
expressed succinctly in \cx,\cxa. 
\item 
~vi)
{\it Hypergeometric Structure:} the kinetic terms have poles only as $X=0$
(respectively $Y=0$), the same positions as the simple poles in the potential. This gives the equation its hypergeometric character.  
\medskip

\noindent
The wave equation for the full metric, without subtraction in the conformal 
factor, was presented in \dd (following \CveticUW). 
Simplifying properties i), ii), iii) remain; but there are additional terms that obstruct properties iv), v), vi). 

The near horizon limit of near extreme Kerr is such that the
wave equation enjoys all the simplifying properties above, except spherical symmetry iv) \BredbergPV,\CveticJN. 
Several other near horizon limits that may be applied to the wave 
equation have the effect of restoring all the simplifying properties \CveticUW. 

\appendix{B}{Derivation of the 6D Lift}
As we have emphasized, the scalar wave equation is separable into radial and 
angular equations. We want to exhibit this factorization explicitly in the geometry.

The form \wf\ of the geometry is a good starting point. The first two terms depend just 
on the radial/temporal variables $x$, $t$, $q$, $S$. The next two terms depend just on the 
angular variables $y$, $\chi$, $\sigma$, $p$, $T$, up to a shift of $\chi$, $\sigma$ 
that is linear in $t$. This latter complication indicates that the angular space is fibered 
over the radial/temporal one, a feature that is well-known from the description of near extreme
rotating black holes \CveticXH. 

It is thus the last term in \wf\ that encodes the coupling between the radial and angular 
parts of the geometry. We can simplify this coupling greatly by lifting to the 6D geometry
\eqn\xa{\eqalign{
\ell^{-2} ds^2_6 & = \Delta ({1\over\mu} d\alpha + {\cal B})^2 + \Delta^{-1/3}ds^2_5~,
}}
where 
\eqn\xb{
{\cal B} = {\mu\over \Delta} (p+q)~.
}
The 6D KK gauge field \xb\ was designed to factorize the offending term in the geometry,
which now takes the form
\eqn\xc{\eqalign{
\ell^{-2}ds^2_6 &= {dx^2\over 4X}   - {X\over \mu^2 S} dt^2  + S(d\alpha - {1\over S} q)^2 \cr
& + {dy^2\over 4Y}+ {Y\over T}
(\Pi_c d\sigma - {1\over\mu}dt- \Pi_s d\chi)^2 + T (d\alpha + {1\over T}p)^2~.
}}
This is what we want: the first line depends just on the 
radial/temporal variables $x$, $t$, $\alpha$, $q$, $S$. 
The second line depends just on the angular variables $y$, $\chi$, $\sigma$, $p$, $T$, up to a 
shift of $\chi$, $\sigma$ that is linear in $t$, $\alpha$. In this representation the angular space is
{\it only} coupled to the radial one by the non-trivial fibration. 

The angular terms can be simplified  so that the  full metric becomes
\eqn\xd{\eqalign{
\ell^{-2}ds^2_6 &= {dx^2\over 4X}   - {X\over \mu^2 S} dt^2  + S(d\alpha - {1\over S} q)^2  + 
{dy^2\over 4Y}+ {Y\over y}d\tilde{\sigma}^2 + y ( d\tilde{\chi} - {ab\over y}d\tilde{\sigma})^2~,
}}
where the shifted coordinates
\eqn\xe{\eqalign{
d\tilde{\chi} &= d\chi - {\Pi_s\over \mu (\Pi^2_c - \Pi^2_s)}dt  + 
\Pi_c d\alpha~,\cr
d\tilde{\sigma} &= d\sigma - {\Pi_c\over \mu (\Pi^2_c - \Pi^2_s)}dt + \Pi_s
d\alpha~.
}}
In this form the metric is recognized as BTZ$\times S^3$. Thus our 
construction associates a locally AdS$_3\times S^3$
geometry to the general black holes we study. 

A conservative interpretation of the auxiliary coordinate $\alpha$ is that it represents a redundancy 
that we can introduce without loss of generality. Certainly wave functions that are independent
of $\alpha$ satisfy the same wave equation in the 6D geometry \xc\ and in the 5D geometry
\wf. From this point of view we have introduced a redundancy in the description,  in order to 
furbish a linear realization of the symmetries. 
 
A more ambitious (and speculative) interpretation of the coordinate $\alpha$ is to identify
it with the physical coordinate along an effective string. In this view the microscopic interpretation 
of black holes involves dependence on $\alpha$ in an essential manner, with the macroscopic
(thermodynamic) description sensitive only to features that are independent of $\alpha$.

\listrefs
\end